\documentclass[a4,twocolumn,aps,pre,showpacs,preprintnumbers,amsmath,amssymb]{revtex4}
\usepackage{epsfig}
\usepackage{graphics}
\usepackage{lipsum}
\usepackage{sidecap}
\usepackage{graphicx} 
\usepackage{dcolumn} 
\usepackage{bm}
\usepackage{sidecap}

\begin{document}
\preprint{APS/123-QED}
\widetext
\title{Observation and characterization of chimera states in coupled dynamical systems with nonlocal coupling}

\author{R.~Gopal$^{1}$}
\author{V.~K.~Chandrasekar$^{1}$}
\author{A.~Venkatesan$^{2}$}
\author{M.~Lakshmanan$^1$}
 
\affiliation{
$^1$Centre for Nonlinear Dynamics, School of Physics, Bharathidasan University, Tiruchirapalli-620024, India\\
$^2$Department of Physics, Nehru Memorial College, Puthanampatti,
Tiruchirapalli 621 007, India.
}

\date{\today}
            
\begin{abstract}
By developing the concepts of strength of incoherence and discontinuity measure, we show that a distinct quantitative characterization of chimera and multichimera states which occur in networks of coupled nonlinear dynamical systems admitting nonlocal interactions of finite radius can be made. These measures also clearly distinguish between chimera/multichimera states (both stable and breathing types) and coherent and incoherent as well as cluster states. The measures provide a straight forward and precise characterization of the various dynamical states in coupled chaotic dynamical  systems irrespective of the complexity of the underlying attractors.
\end{abstract}
\maketitle

\section{\label{sec:intro}Introduction}

Chimera states are highly counter intuitive structures coexisting with coherent (synchronized) and incoherent (desynchronized) oscillations, originally identified in populations of nonlocally coupled oscillators such as  Ginzburg-Landau systems, R\"ossler oscillators and logistic maps~\cite{kura1995,kura1996,kura1998,kura2000} and in recent times in identical phase oscillators~\cite{kuramoto2002, shima2004,abrams2004,abrams12006,martens2010}. They have been further studied extensively in various generalized situations, including time delay. These include networks occurring in neuroscience~\cite{battagila2007, vicente2008}, Josephson junction arrays~\cite{wiesenfield1996}, and electrochemical systems~\cite{mazouz1997}. Real world examples include uni-hemispheric sleep in certain animals~\cite{rottenberg2000}, where the awake side of the brain shows desynchronized  electrical activity, while the sleeping side is highly synchronized~\cite{motter2010}.

In recent times several theoretical studies~\cite{kaw2007,abrams2008,hsheeba2010,martens12010,bordyugov2010} and experimental investigations~\cite{hagers2012,tinsley2012,nkoma2013,martens2013,larger2013} have established chimera as a robust concept  occurring in varied complex networks, including chaotic dynamical systems~\cite{omel2011,omel2012,hagers2012}. For example networks of chaotic dynamical systems with nonlocal coupling exhibit coexisting spatial domains of coherence and incoherence ~\cite{omel2011,omel2012}, coherent travelling waves ~\cite{dziubak2013} and spiral wave chimera states~\cite{gu2013}. Experimentally they have been observed in optical~\cite{hagers2012} and chemical systems~\cite{tinsley2012,nkoma2013}, a mechanical experiment consisting of two populations of metronomes~\cite{martens2013} and a modified Ikeda time delay circuit system~\cite{larger2013}. It has been recently pointed out~\cite{omel2011,omel2012} that transition from spatially coherent to incoherent state occurs via chimera state in models of coupled logistic map and R\"ossler systems with nonlocal interaction. A very  interesting observation in this connection is the identification of chimera and multichimera states (two or more incoherent domains) in Fitzhugh-Nagumo oscillators, and their characterization by using the value of mean phase velocity (frequency)~\cite{omel2013}. One may note that the  notion of phase velocity essentially requires periodic behavior of individual oscillators, while this is difficult to extend to chaotic systems. 

Under suitable circumstances it has been shown that breathing chimera states can also arise, for example in two populations of phase oscillators~\cite{abrams2008,hsheeba2010}  and networks of Lorenz systems as we point out later in this paper where cluster states can also arise. Thus there arises an urgent need to characterize the transition from incoherent to coherent state via chimera (or breathing chimera) and  multichimera  states in terms of definitive quantitative measures.  We successfully address this problem by developing suitable statistical measures using the time series of the networks in terms of measures designated as strength of incoherence($SI$) and discontinuity measure($DM$) deducible from a local standard deviation analysis. A clear quantification of chimera and multichimera states is given in terms of nonzero (but less than unity) values of $SI$ and   positive integer values of $DM$. The coherent state is characterized by zero $SI$ and $DM$ values, while an incoherent state takes unit $SI$ value and zero $DM$ value. Breathing chimera and cluster states can also be characterized appropriately. 

In this connection we also wish to point out that Kuramoto and his co-workers~\cite{kura1995,kura1996,kura1998,kura2000} in their early works  have characterized the various spatio-temporal patterns  occurring in nonlocally coupled systems as a function of coupling strength by a spatial correlation function for the difference of the field variables. They have shown that the correlation function  exhibits a power law dependence on the distance and a discontinuous peak at the origin  when the coupling constant is decreased below some critical value, indicating that the spatial pattern is statistically discontinuous, which was also explained through a stochastic model and more generalized multifractal analysis. It was also shown that the correlations and fluctuations obey a power law similar to the one in the fully developed Navier- Stokes turbulence except that the exponent changes continuously with the coupling strength. However, it is not clear from these studies that how quantitative characterization to distinguish different dynamical states such as coherent, chimera, multichimera, including stationary and breathing type, cluster and incoherent states can be identified. We believe that our present investigations gives suitable clear quantitative measures to distinguish these various states.

Our characterization works for systems admitting both phase coherent and non-phase coherent attractors where a principal frequency cannot be easily identified. Our findings also show that even without introducing the concepts of phase and frequency one can succeed to distinguish different dynamical states, namely coherent, incoherent, chimera, multichimera and cluster states in coupled dynamical systems.

This paper is structured as follows. In Section II we introduce the quantitative measures, namely the strength of incoherence $(SI)$ and discontinuity measure $(DM)$ as statistical tools to quantify the different dynamical states. In Sections III, we present our analysis of various dynamical states, including coherent, chimera, incoherent and cluster states in four different networks of nonlinear dynamical systems, and quantitative characterization of these states and their transitions in terms of the new statistical measures.  The paper concludes with a summary in Sec. IV. In Appendix A we describe the difficulty in characterizing different dynamical states by using the original state variables and the necessity to introduce transformed variables. In Appendix B we indicate the method of identification of cluster states. 

\begin{figure}
\centering
\includegraphics[width=1.00\columnwidth=1.00]{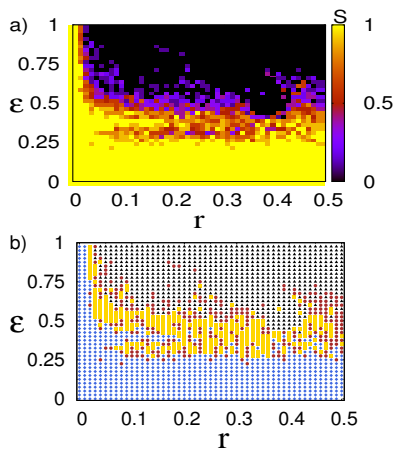}
\caption{\label{fig1}(color online) Two parameter ($r,\epsilon$) phase diagrams for $N=100$ 
coupled Mackey-Glass time delay systems. (a) Strength of incoherence $S$: This figure indicates regions of coherence (black), incoherence (yellow/white),  chimera and multichimera states (red/blue/gray). (b) Discontinuity measure($\eta$): This figure indicates regions of coherence  (black, $\blacktriangle$),incoherence (blue, $\diamond$), chimera (brown, $\bullet$) and multichimera states (gold, {\tiny{$\blacksquare$}}). The system parameters are $\alpha=1.00$, $\beta=2.00$, $\tau=2.00$ (individual nodes are evolving chaotically).}
\end{figure}

\begin{figure}
\centering
\includegraphics[width=1.00\columnwidth=1.00]{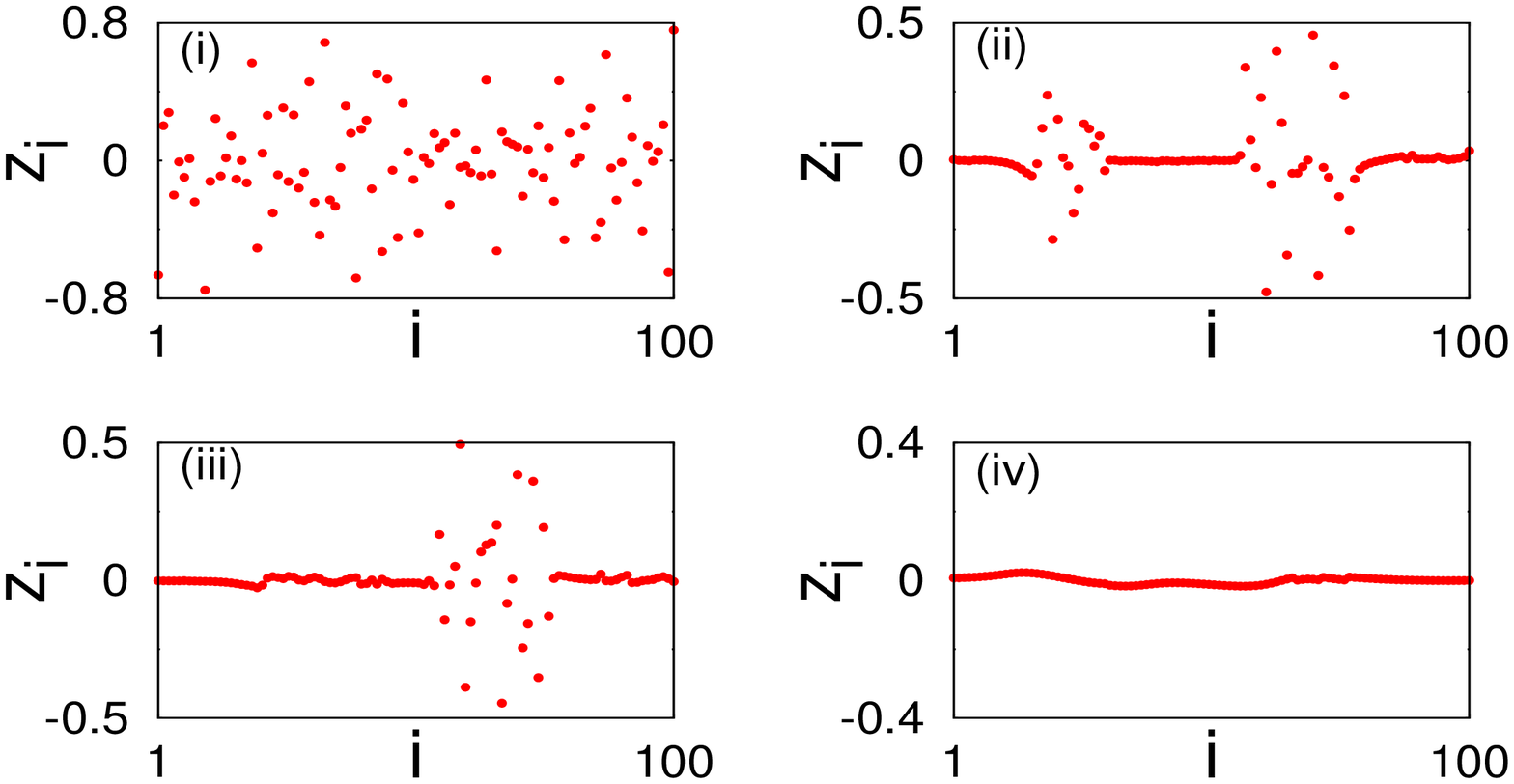}
\caption{\label{fig2}(color online) Snapshots of nonlocally coupled MG time delay system for different values of coupling strength in terms of new state variable $z_{i}$: (i) incoherent state, $\epsilon=0.15$ (ii)  multichimera  state, $\epsilon=0.50$ (iii) chimera state,  $\epsilon=0.56$ and (iv) coherent state, $\epsilon=0.75$. The coupling radius is fixed at $r=0.3$, where  $N=100$. Other parameters are as in Fig.1.}
\end{figure}

\section{Developing Quantitative Measures}

In order to develop the quantitative measures , we consider a network of coupled dynamical systems with nonlocal interactions of finite radius represented by 
\begin{equation}
{\bf{\dot{x}}}_{i}={\bf{F}}_{i}({\bf{x}}_{i},{\bf{x}}_{i,\tau})+\frac{\boldsymbol{\bar{\epsilon}}}{2P}\sum_{j=i-P}^{j=i+P}(\bf{x}_{j}-\bf{x}_{i})
\end{equation}
where $i=1,2,....N$, ${\bf{{x}}}_{i}={\bf{{x}}}_{i}(t)=[x_{1,i}, x_{2,i},....x_{d,i}]^{T} \in 
\mathbb{R}^{d}$, ${\bf{x}}_{i,\tau}={\bf{x}}_{i}(t-\tau)$, $\tau$: constant, and ${\bf{F}}_{i}({\bf{x}}_{i}, {\bf{x}}_{i,\tau})=[{F}_{1}({\bf{x}}_{i}, {\bf{x}}_{i,\tau}), {F}_{2}({\bf{x}}_{i}, {\bf{x}}_{i,\tau}), ...., {F}_{d}({\bf{x}}_{i}, {\bf{x}}_{i,\tau})]^{T}$. Thus ${\bf{{x}}}_{i}(t)$ represents the state vector of the $i^{th}$ oscillator.
In (1) $\boldsymbol{\bar {\epsilon}}$  denotes the coupling matrix and  $P$  specifies the number of neighbors in each direction on a ring so that the coupling radius $r=P/N$. In this paper, we consider  systems having different kinds of attractors, namely (i) non-phase coherent attractors: (a) Mackey-Glass (delay) system and (b) Lorenz (nondelay) system, and (ii) phase coherent attractors: (c) R\"ossler (chaotic) system and (d)  Fitzhugh-Nagumo (periodic) oscillators.

\begin{figure}
\centering
\includegraphics[width=1.00\columnwidth=1.00]{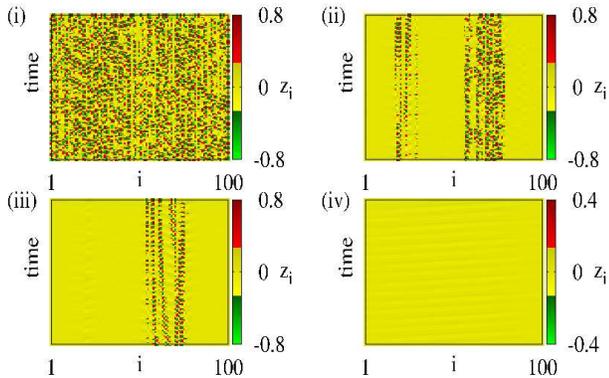}
\caption{\label{fig3}(color online) Space-time plots of nonlocally coupled MG time delay system for different values of coupling strength in terms of new state variable $z_{i}$: (i) incoherent state, $\epsilon=0.15$ (ii)  multichimera  state, $\epsilon=0.50$ (iii) chimera state,  $\epsilon=0.56$ (iv) coherent state, $\epsilon=0.75$. The coupling radius is fixed at $r=0.3$, where  $N=100$. Other parameters are as in Fig.1}
\end{figure}

\begin{figure}
\centering
\includegraphics[width=1.00\columnwidth=1.00]{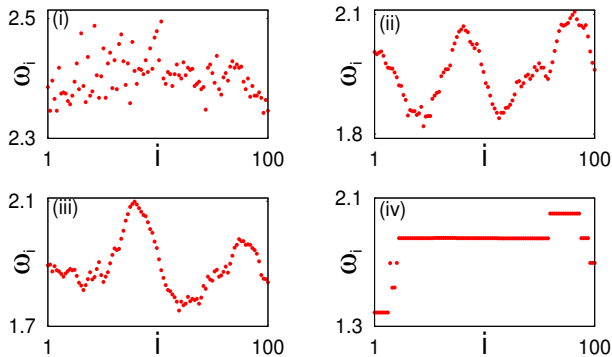}
\caption{\label{fig4}(Color online) Mean phase velocities (frequency) $\omega_{i}$ of  MG time delay system  corresponding to fig. 3: (i) incoherent state, $\epsilon=0.15$ (ii)  multichimera  state, $\epsilon=0.50$ (iii) chimera state,  $\epsilon=0.56$ (iv) coherent state, $\epsilon=0.75$. The coupling radius is fixed at $r=0.3$, where  $N=100$. Other parameters are as in Fig.1.}
\end{figure}

To develop suitable quantitative measures, we note that the original dynamical variables ${\bf{x}}_{i}$ are not the most appropriate ones (see Appendix A). So we introduce a transformation of the state variables ${\bf{x}}_{i}$ to new variables ${\bf{z}}_{i}$, $i=1,2,...,N$, where ${\bf{z}}_{i}={\bf{x}}_{i}-{\bf{x}}_{i+1}$, ${\bf{z}}_{i}=(z_{1,i},z_{2,i},...z_{d,i}) \in \mathbb{R}^{d}$. Now, the occurrence of different synchronized states in the coupled system (1) can be better illustrated  using the new state variables ${\bf{z}}_{i}$. We also note that when the $i^{th}$ and $i+1^{th}$ oscillators are oscillating coherently the value of $z_{l,i}$ is minimum (To be precise $z_{l,i}\rightarrow 0$ as $N\rightarrow\infty$).

 On the other hand when two neighboring oscillators $i$ and $i+1$ are oscillating  incoherently, $z_{l,i}$'s take  values between $\pm|x_{l,i,max}-x_{l,i,min}|$ (where $x_{l,i,max/min}$ are upper/lower bounds of the allowed values of $x_{l,i}$'s ). Thus in the case of coherent states all the $z_{l,i}$'s  take a minimum value for all times, while in the case of an  incoherent state  $z_{l,i}$'s  get distributed between $\pm|x_{l,i,max}-x_{l,i,min}|$. However, in the chimera state some of the $z_{l,i}$'s may take the same value while the others may be distributed  over the above range. In order to quantify the synchronized states clearly, we introduce the notion of standard deviation for the asymptotic state as
\begin{equation}
\sigma_{l}=\Big<\noindent \sqrt{\frac{1}{N}\sum_{i=1}^{N}[z_{l,i}-<z_{l}>]^2} \hspace{0.1cm} \Big>_{t},
\end{equation}
where $z_{l,i}=x_{l,i}-x_{l,i+1}$, $l=1,2...d$, $i=1,2...N$, and  $<z_{l}>=\frac{1}{N}\sum_{i=1}^{N}z_{l,i}(t)$ and $\langle ...\rangle_{t}$ denotes average over time. Consequently $\sigma_{l}$'s take a value zero for coherent states and nonzero values for both incoherent and chimera states. We also note that one is unable to distinguish between incoherent and chimera states using $\sigma_{l}$ alone because in both the cases $\sigma_{l}$ can take similar non-zero values.  To overcome this difficulty we divide the oscillators into $M$ (even) bins of equal length $n=N/M$. Consequently, we introduce the local standard deviation $\sigma_{l}(m)$ which can be defined as
\begin{equation}
\sigma_{l}(m)=\Big<\noindent \sqrt{\frac{1}{n}\sum_{j=n(m-1)+1}^{mn}[z_{l,j}-<z_{l}>]^2} \hspace{0.1cm}\Big>_{t}, \hspace{0.1cm}m=1,2,...M.
\end{equation}

The above quantity $\sigma_{l}(m)$  is calculated for every successive $n$ number of oscillators. Using (3) we can introduce a $SI$ as \\

\begin{equation} 
S=1-\frac{\sum_{m=1}^{M}s_{m}}{M}, \hspace{0.1cm}  s_{m}=\Theta(\delta-\sigma_{l}(m)),
\end{equation}
where $\Theta(.)$ is the Heaviside step function, and $\delta$ is a predefined threshold that is reasonably small. Here, we take $\delta$ as a certain percentage value of difference between $x_{l,i,max}$ and $x_{l,i,min}$.  Thus when $\sigma_{l}(m)$ is less than $\delta$, the value of $s_{m}=1$, otherwise it is '0'. Consequently, $SI$  takes the values $S=1$ or $S=0$ or  $0 < S < 1$ for incoherent, coherent and chimera /multichimera  states, respectively.

To gain a better understanding and to distinguish further between chimera and multichimera states, we also introduce a $DM$, based on the distribution of $s_{m}$ in (4). It is defined as 

\begin{equation}
\eta=\frac{\sum_{i=1}^{M} |s_{i}-s_{i+1}|}{2}, \hspace{0.3cm} (s_{M+1}=s_{1})
\end{equation}
In this case $\eta$ takes a value '1' for chimera state,  and positive integer value greater than '1' for multichimera states. For breathing and cluster states see below. The characterization is summarized in Table - I.

\begin{table}
Table - I
\begin{ruledtabular}
\begin{tabular}{lcr}
Dynamical state & ($S$, $\eta$) & Remarks\\
\hline
   coherent &		(0, 0) &\\  
   chimera  &		(c, 1) & $0<c<1$\\ 
   multichimera &	(c, d)  & $2\leq d \leq M/2$\\ 
   incoherent & 	(1, 0)  \\  
\end{tabular}
\end{ruledtabular}
\end{table}  
 
\section{Characterization of different dynamical states and their transitions in coupled dynamical systems}
In this section we will apply the above quantitative criteria to characterize the different dynamical states and their transitions in coupled nonlinear dynamical systems with nonlocal interactions. We consider four specific models in our study as described below.
\subsection{Mackey-Glass system}

 To demonstrate the above characterization,  we first consider the Mackey- Glass (MG) system~\cite{mackey1977,lakshmanan2011}  ${\bf{F}}({\bf{x},x_{\tau}})=-\alpha x+\frac{\beta x(t-\tau)}{[1+x(t-\tau)^{10}]}$ in (1) with parameters  chosen as $\alpha=1$, $\beta=2$, $\tau=2$ (so that individual nodes oscillate chaotically in the absence of coupling). In this case the coupling matrix becomes a scalar, $\boldsymbol{\bar {\epsilon}}={\epsilon}$.  Let us first discuss the distribution of synchronized states in the $(r,\epsilon)$ parameter space admitted by the MG equation with nonlocal coupling. Fig. 1(a) shows the two parameter phase diagram of MG system  when the individual nodes are oscillating chaotically and incoherently in the absence of coupling.  On introducing the interaction with a coupling radius $r$ and coupling strength $\epsilon$ the incoherent state persists up to certain values of coupling strength $\epsilon=\epsilon_{i}$. This is clearly seen from Fig. 1(a) where $S=1$ and is denoted by the yellow (white) region. When $\epsilon_{i}<\epsilon<\epsilon_{c}$, we find that chimera / multichimera states exist, where the values of $S$ varies between 0 and 1 (gray region in Fig. 1(a)). Upon increasing the value of $\epsilon$ beyond $\epsilon_{c}$, the chimera / multichimera state loses its stability and transits into a coherent state. In this state the value of $S$ is zero and is marked as black in Fig. 1(a). 

We further present the two parameter phase diagram in terms of $\eta$ as shown in Fig. 1(b). This figure clearly indicates regions of coherence (black, $\blacktriangle$), incoherence (blue, $\diamond$), chimera (brown, $\bullet$) and multichimera states (gold, {\tiny{$\blacksquare$}}). The measure $\eta$ takes a value zero for coherent/incoherent state, '1' for chimera, and an integer value greater than '1' ($2\leq \eta \leq M/2 $) for multichimera states.

\begin{figure*}
\centering
\includegraphics[width=1.7\columnwidth=1.10]{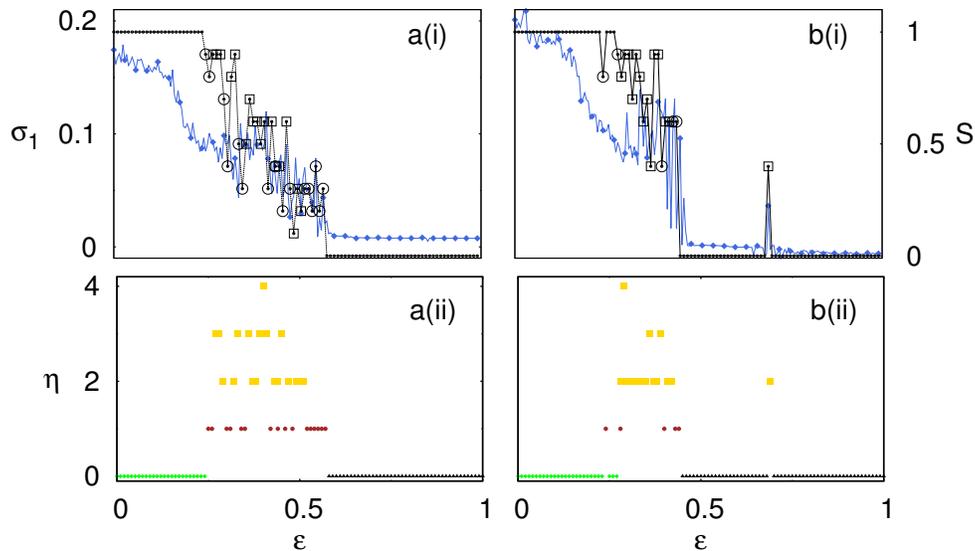}
\caption{\label{fig5} (color online) (i) Standard deviation $\sigma_{1}$ (blue curve) , strength of incoherence $S$ (black curve)  and (ii) discontinuity measure $\eta$  versus coupling strength $\epsilon$ for nonlocal interaction of coupled MG time delay system  with parameters $\alpha=1.00$, $\beta=2.00$ and $N=100$; coupling radius is taken as (a) $r=0.3$ (b) $r=0.4$. Black squares and circles in the $S$ plot indicate multichimera and chimera states, respectively.} 
\end{figure*}

Next, we consider the transition of chimera and multichimera states as well as  coherent and incoherent states in the coupled  MG equations. In the set of Figs. 2, we fix the coupling radius at $r=0.3$ and vary the  coupling strength $\epsilon$.  We display the typical scenario of different synchronized states, namely incoherent (Fig. 2(i)), multichimera (Fig. 2(ii)), chimera (Fig. 2(iii)) and coherent states (Fig. 2(iv)), which are snapshots of  the transformed  state variables $z_{i}$ (Here  ${\bf{z}}_{i}$ is a scalar). Corresponding space-time plots are shown in Fig. 3. Here, during the transition from an incoherent to a  coherent state, we find a multichimera state besides the chimera state.

As may be seen from these figures, initially the system is in an incoherent state for $\epsilon=0.15$ (Figs. 2(i) and 3(i)). In this state the values of $z_{i}$'s are randomly distributed. On increasing the coupling strength to $\epsilon=0.50$ we find the occurrence of multichimera state (Figs. 2(ii) and 3(ii)). In this state two groups of oscillators are evolving in an incoherent manner (oscillator indices 20-30 and 60-80) between coherent oscillators. On further increasing the coupling strength to  $\epsilon=0.56$ (Figs. 2(iii) and 3(iii)), we find  a single chimera state. In this state the oscillators with indices 60 to 80 are in a desynchronized state while the remaining oscillators are in a coherent or spatially synchronized state. For $\epsilon=0.75$, the system enters into a single coherent state (Figs. 2(iv) and 3(iv)). In this state the values of $z_{i}$'s approach a minimum for all times. Figures 4(i) - 4(iv) indicate mean phase velocities $\omega_{i}$ (frequency)~\cite{omel2013}  corresponding to incoherent, multichimera, chimera and coherent states, respectively. The values of $\omega_{i}$ for each oscillator is calculated as $\omega_{i}=2\pi M_{i}/\Delta T$, $i=1,2,3...N$, where $M_{i}$ is the number of maxima of the  time series $x_{i}(t)$ of the $i^{th}$ oscillator during the time interval $\Delta T$. Note that the distribution of  $\omega_{i}$ fails to distinguish different states (particularly chimera and multichimera states).

\begin{figure}
\centering
\includegraphics[width=1.00\columnwidth=1.00]{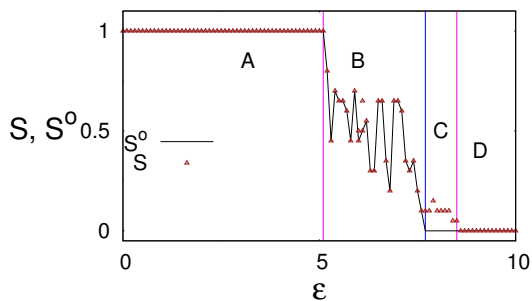}
\caption{\label{fig6}(color online) Strength of incoherence $S$ and its modified form $S^{o}$ (after removal of deviated points) as a function of coupling strength ($\epsilon$) for nonlocally coupled Lorenz systems with coupling radius $r=0.3$ and $N=500$. Here, incoherent (A), chimera/multichimera states (B), cluster states (C) and coherent state (D) are identified.} 
\end{figure}

In Figs. 5a(i) and b(i) we demonstrate the behavior of standard deviation $\sigma_{1}$ (red/gray) and the strength of incoherence  $S$ (black) as a function of the coupling strength $\epsilon$ for two different values of coupling radius, $r=0.3$ and $r=0.4$. The distribution of $\sigma_{1}$ (Eq.(2)) is shown in Fig. 5a(i) which indicates that it takes nonzero values for  incoherent and chimera states.  As $\epsilon$  increases, the value of $\sigma_{1}$  approaches zero for $\epsilon>\epsilon_{c}$, where the coherent state occurs. The same behavior is also observed for the coupling radius $r=0.4$ as shown in Fig. 5b(i). Thus by using the standard deviation $\sigma_{1}$ we are able to distinguish between  the incoherent/chimera and coherent states only, while it does not distinguish clearly the chimera state from incoherent state. In order to distinguish these two states clearly we plot the  $SI$($S$) in Figs. 5a(i) and 5b(i) corresponding to the above values of $\sigma_{1}$. Here, $S$ takes a unit value for the incoherent state, while it takes a value zero  for the coherent state. On the other hand $S$ oscillates between 0 and 1 for both chimera and multichimera states. To distinguish the last two states  we also plot the $DM$ ($\eta$) in terms of $\epsilon$ in Figs. 5a(ii) and 5b(ii). We find that  $\eta$ takes a value unity for chimera states, and  a higher integer value for multichimera states. 

\begin{figure}
\centering
\includegraphics[width=1.00\columnwidth=1.0]{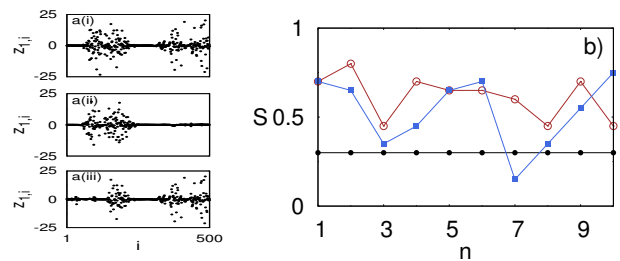}
\caption{\label{fig7}(color online) Characterization of breathing chimera state: (a) Snapshots of the coupled Lorenz system illustrates breathing chimera state ($\epsilon=5.60$) at (i) $t=2800$, (ii) $t=2900$ and (iii) $t=2990$.  (b) Strength of incoherence $S$ as a function of ${n}$ for the values of $\epsilon$ in MG equation for $\epsilon=0.5$ (dots) and Lorenz systems (open circles ($\epsilon=5.60$) and black squares ($\epsilon=6.20$))} 
\end{figure}

\begin{figure}
\centering
\includegraphics[width=1.00\columnwidth=0.80]{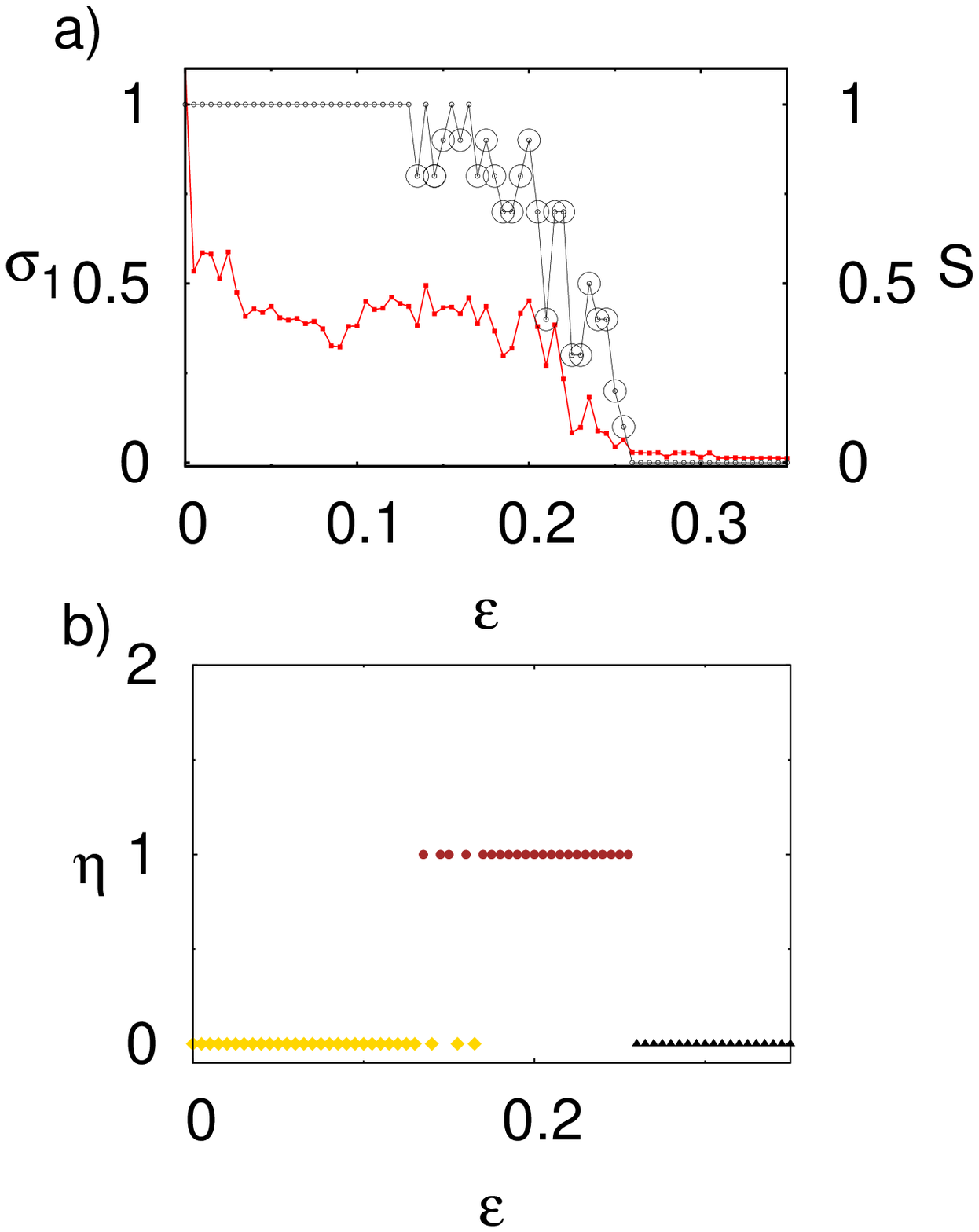}
\caption{\label{fig8}(color online) (a) Standard deviation $\sigma_{1}$ (red) and strength of incoherence $S$ (black), and (b) discontinuity measure ($\eta$) versus coupling strength $\epsilon$ for non-local interaction of R\"ossler systems with $N=100$, $r=0.35$} 
\end{figure}

\subsection{Lorenz system} 
To test the universality of  characterization by $S$ and $\eta$,  we next investigate a system of nonlocally coupled Lorenz oscillators with  ${\bf{F(x)}}=[\sigma(x_{2}-x_{1}),  x_{1}(\rho-x_{3})-x_{2}, x_{1}x_{2}-\beta x_{3}]^{T}$, where the diagonal elements of $\boldsymbol{\bar {\epsilon}}$ are nonzero ($\epsilon$) in Eq. (1). Here, the node parameters are fixed at the values $\sigma=10$, $\rho=28$, $\beta=8/3$ and $\tau=0$. We fix $r=0.3$ and vary the value of $\epsilon$. Identification of chimera/multichimera states along with incoherent, coherent and cluster states in terms of $S$  plots is made in Fig. 6. In this case, initially the system is in an incoherent state (A) up to $\epsilon \approx 5.10$, and increasing the value of it the system exhibits chimera/multichimera (B) states. On further increasing the value of $\epsilon$, the system transits into a coherent state (D). During this transition one can also observe cluster states (See Appendix B), which are two (or more) independent groups of coherent states, marked C in Fig. 6. The cluster states correspond to a finite number of distinct coherent profiles, having finite discontinuity in the  ${\bf{x}}_{i}$ variables. In the transformed ${\bf{z}}_{i}$ variables the profile will be  essentially a continuous curve with 2$q$ distinct deviating points corresponding to the $q$ discontinuous cluster profiles in the ${\bf{x}}_{i}$ variables. These discrete points are removed by the method of removable discontinuity~\cite{dis1992}. In Fig. 6 we have plotted $S$ (before removing deviated points of ${\bf{z}}_{i}$) and $S^{o}$ (after removing deviated points of ${\bf{z}}_{i}$) as a function of $\epsilon$. It indicates that the values of $S$ and $S^{o}$ remains the same in the case of coherent, incoherent and chimera/multichimera states. In the case of cluster states $S$ takes a nonzero value, but $S^{o}$ takes a zero value (which is marked as C in Fig. 6).

Interestingly, the occurrence of chimera and multichimera states in the coupled Lorenz systems are of breathing type compared to the stable chimera states identified in the case of coupled  MG equations (over a time $T$). The occurrence of breathing chimera state is demonstrated in Fig. 7(a), where we present snapshots of the variables $z_{1,i}$ for three different times  (i) $t=2800$ (ii) $t=2900$ (iii) $t=2990$ for $\epsilon=5.60$ showing the breathing nature of the chimera (which gets repeated in $t$). 

In order to distinguish such breathing  chimera states compared to stable chimera states, we carried out the following analysis instead of characterization through $DM$. The total time $t\in(0,T)$ of spatiotemporal evolution is also divided into $k$ bins ($t_{n}, n=1,2,...k$). Now, each bin has $t_{s}$ time units ($t_{s}=T/k$). The calculation of $S$  using Eq. (4) for each bin can be performed as before and it gives rise to $k$ number of $S$ values. The $S$ values for stable chimera state of MG equations for $\epsilon=0.50$ and breathing chimera states of Lorenz system for $\epsilon=5.60$ and $\epsilon=6.20$ are shown in Fig. 7(b). The figure clearly indicates  that the value of $S$ remains constant for the stable chimera state (MG system) and varies for breathing chimera states (Lorenz system) as a function of $n$.

\begin{figure*}
\centering
\includegraphics[width=1.5\columnwidth=1.5]{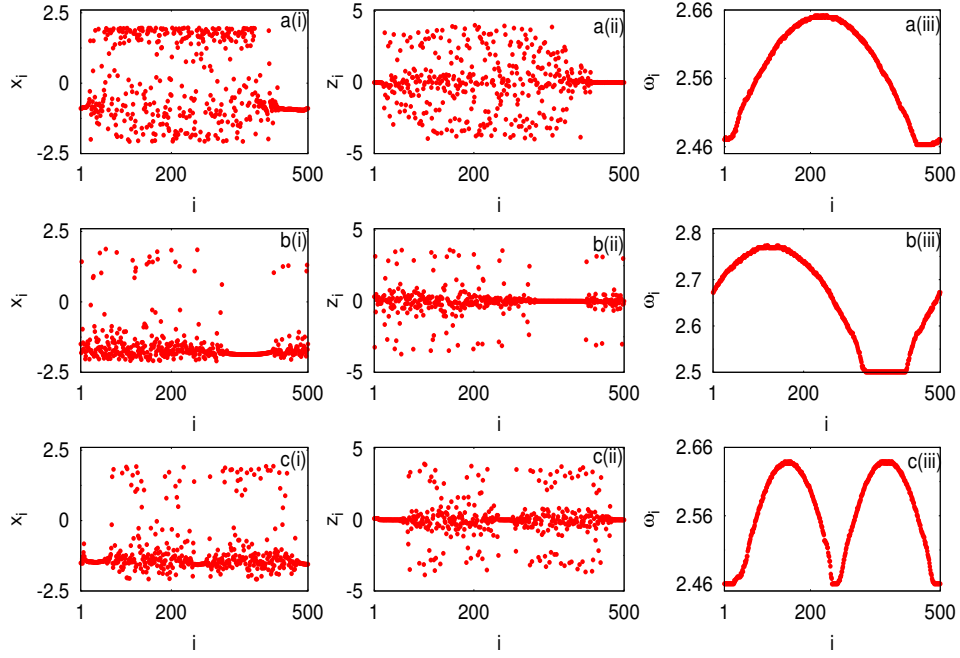}
\caption{\label{fig9}(color online)(i) Snapshots of variables $x_{i}$, (ii) Snapshots of variables $z_{i}$ and (iii) mean phase velocities $\omega_{i}$ (frequency). (a) chimera state with one incoherent domain, $\epsilon=0.20$  (b) chimera state with one incoherent domain, $\epsilon=0.28$  (c) multichimera state with two incoherent domains, $\epsilon=0.32$   for nonlocal interaction of coupled FHN oscillators with  $N=500$ and coupling radius $r=0.33$.}
\end{figure*}

\begin{figure}
\centering
\includegraphics[width=1.0\columnwidth=1.20]{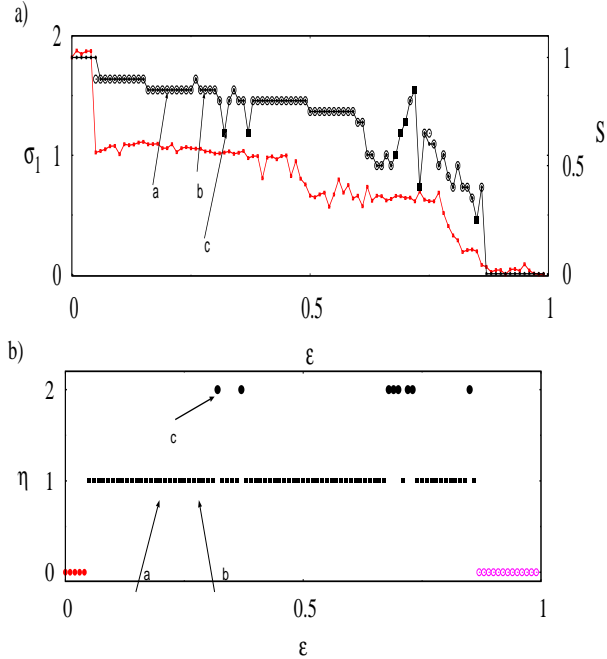}
\caption{\label{fig10} (color online)(a) Standard deviation $\sigma_{1}$ (red/gray) and strength of incoherence $S$ (black),  and (b) discontinuity measure $\eta$  versus coupling strength $\epsilon$. Black squares and circles in the $S$ plot indicate multichimera and chimera states respectively. The  parameters chosen are as in Fig. 9. Points a, b and c correspond to the three cases of Fig. 9.} 
\end{figure}

\subsection{R\"ossler system}
Next we consider the  incoherent-coherent transition via chimera state, in  a system of nonlocally coupled R\"ossler systems. Here, we choose $F(\mathbf{x})=[ -x_{2}-x_{3},  x_{1}+ax_{2}, b+x_{3}(x_{1}-c)]^{T}$ where the diagonal elements of $\boldsymbol{\bar {\epsilon}}$ are nonzero ($\epsilon$) in Eq.(1). The system parameters are chosen as $a=0.42$, $b=2$ and $c=4$ showing chaotic dynamics in the uncoupled case. 
The present study clearly distinguishes the  various dynamical states namely incoherent, chimera, and coherent states, through the quantities $S$ and $\eta$. Fig. 8 presents the values of $\sigma_{1}$, $S$, and $\eta$ for the coupled R\"ossler system  as a function of $\epsilon$.

\subsection{FitzHugh-Nagumo (FHN) oscillator}
Now, we consider a ring of $N$ nonlocally coupled FHN oscillators given by 
\begin{eqnarray}
b \frac{dx_{i}}{dt}=x_{i}-\frac{x_{i}^{3}}{3}-y_{i}+\frac{\epsilon}{2P}\sum_{j=i-P}^{j=i+P}[b_{xx}(x_{j}-x_{i}) \notag \\+b_{xy}(y_{j}-y_{i})],   \notag \\
\frac{dy_{i}}{dt}=x_{i}+a+\frac{\epsilon}{2P}\sum_{j=i-P}^{j=i+P}[b_{yx}(x_{j}-x_{i}) \notag \\
+b_{yy}(y_{j}-y_{i})], \notag 
\end{eqnarray}
where $x_{i}$ and $y_{i}$ are the activator and inhibitor variables, $a$ is a threshold parameter and   $b$ is a small parameter characterizing the time scale of separation.  Then, the form of the rotational coupling matrix is 

\begin{equation}
B=\begin{pmatrix} b_{xx} & b_{xy} \\ b_{yx} & b_{yy} \end{pmatrix}=\begin{pmatrix} cos \phi & sin \phi \\ -sin  \phi & cos \phi \end{pmatrix}     \notag 
\end{equation}

depending on a single parameter $\phi \in [-\pi, \pi]$. For the investigation of different states, we fix the parameters as $b=0.05$, $a=0.5$, $\phi=\pi/2-0.1$ and coupling radius as $r=0.33$ and vary the value of $\epsilon$. In Ref.~\cite{omel2013}, the transition from chimera states to multichimera states is studied by using mean phase velocities by varying the values of $\epsilon$ and $r$ (Fig. 9). In the present study chimera and multichimera states are classified by using the values of  $SI$ and $DM$, which clearly identify all the collective states distinctly (see Fig. 10).

 The snapshots of both $x_{i}$ and $z_{i}$ and their corresponding mean phase velocities ($\omega_{i}$) are shown in Fig. 9. Here, Figs. 9(a) and 9(b) illustrate chimera states (which consist of one incoherent and one coherent structures), while Figs. 9(c) demonstrates multichimera states (which consist of two incoherent and two coherent structures) for a ring of $N=500$ non-locally coupled FHN oscillators. A detailed characterization of these states, using the measures $\sigma_{1}$, $S$ and $\eta$, is presented in Figs. 10. 

From Figs. 9 and 10, we note that for the cases of chimera states for $\epsilon=0.20$ and $\epsilon=0.28$ the mean phase velocity consists of one incoherent domain (Figs. 9a(iii),b(iii)). For these two cases, the values of $S$ lie between 0 and 1 while  $\eta$ takes a value 1 (marked a, b in Fig. 10(a) and (b)). In the multichimera state for $\epsilon=0.32$, the mean phase velocity consists of two incoherent domain (Fig. 9c(iii)). Here again $0<S<1$ but $\eta$ takes a value 2 (marked c in Fig. 10(a) and (b)). Therefore, our studies agree with the existing identification of chimera/multichimera  states as a function of mean phase velocities for FHN oscillator system.

\section{Conclusion}
In summary, we have presented  a distinct set of quantitative criteria for chimera and multichimera states in coupled dynamical systems with nonlocal coupling. We have also studied the transition from incoherent  to coherent states via chimera/multichimera states by using strength of incoherence and discontinuity measure. By developing a two parameter phase diagram in terms of these quantifiers we have identified different synchronized states in coupled Mackey-Glass systems and then extended the study to coupled Lorenz systems, coupled Fitzhugh-Nagumo oscillators and coupled R\"ossler systems with nonlocal interaction. These results confirm that the  proposed measures are universally applicable to networks of coupled dynamical systems. 

\begin{acknowledgments}
The work of R.G., V.K.C and M.L has been supported by the Department of Science 
and Technology (DST), Government of India sponsored IRHPA research project. M.L. has also been supported by a DAE Raja Ramanna fellowship and a DST Ramanna program.
\end{acknowledgments}

\begin{appendix}
\section*{Appendix A: Difficulty in characterizing different dynamical states by using original state variables: Mackey-Glass time delay system}
In this Appendix, we first display the typical scenario of transition from incoherent to coherent states in terms of the original state variables $x_{i}$ and point out that a direct statistical analysis of the corresponding data fails to clearly distinguish chimeras from incoherent states. The corresponding snapshots and space-time plots are shown in Figs. 11 and 12 respectively.  In Fig. 11(a) the values of $x_{i}$'s as a function of the oscillator index '$i$' are shown for $\epsilon=0.15$ corresponding to an incoherent state.  On increasing the value to $\epsilon= 0.50$ (Fig. 11(b)) a multichimera state is obtained. Then, at the value  $\epsilon= 0.56$ (Fig. 11(c)) a  chimera state results. Then, at $\epsilon=0.75$, the chimera state loses its stability and transits to a coherent state (Fig. 11(d)).

\begin{figure}
\centering
\includegraphics[width=1.0\columnwidth=1.0]{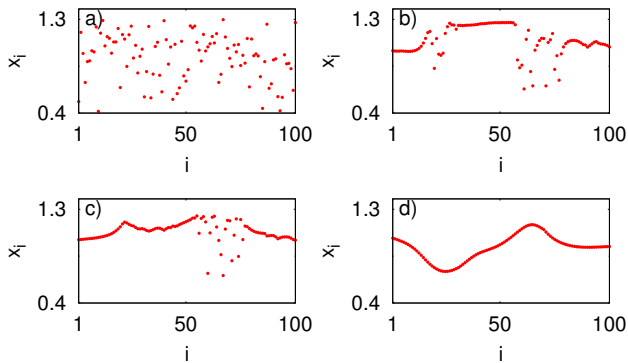}
\caption{\label{fig11}(color online) Snapshots of the variables $x_{i}$ for coupled Mackey-Glass time delay system for a fixed coupling radius $r=0.3$. Occurrence  of  (a) incoherent state for  $\epsilon=0.15$, (b) multichimera state for $\epsilon=0.50$, (c) chimera state for $ \epsilon=0.56$ and (d) coherent state for $\epsilon=0.75$.}
\end{figure}

\begin{figure}
\centering
\includegraphics[width=1.0\columnwidth=1.0]{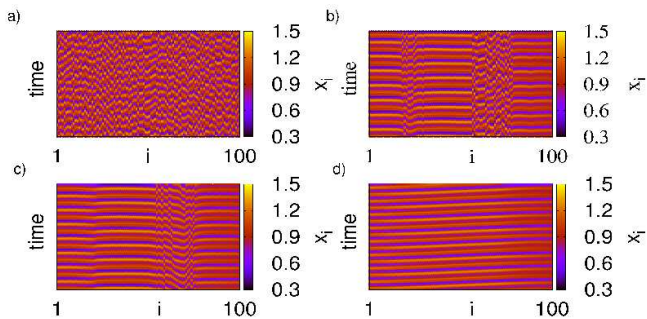}
\caption{\label{fig12}(color online) Space-time plots of $x_{i}$'s for coupled MG time delay system  for fixed coupling radius $r=0.3$. Occurrence  of  (a) incoherent state for $\epsilon=0.15$, (b) multichimera state for $\epsilon=0.50$, (c) chimera state for $ \epsilon=0.56$ and (d) coherent state for $\epsilon=0.75$. The other parameters are fixed as $\alpha=1.00$ $\beta=2$, $\tau=2.00$ and $N=100$ in the  MG equation.}
\end{figure}

\begin{figure}
\centering
\includegraphics[width=1.0\columnwidth=1.0]{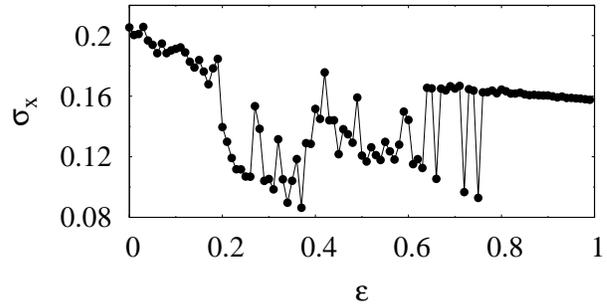}
\caption{\label{fig13}(color online)  Standard deviation $\sigma_{x}$ of the state variables $x_{i}$ versus $\epsilon$ for the coupled MG time delay system. The other parameters are fixed as $\alpha=1.00$, $\beta=2$, $\tau=2$ and $N=100$.}
\end{figure}

The above snapshots of $x_{i}$'s and corresponding space-time plots reveal the following: 1) A random distribution of dynamical variables for incoherent state (Figs. 11(a) and 12(a)). 2) A random distribution of two or more groups of oscillators interspersed by groups coherent oscillators (Figs. 11(b) and 12(b)). 3) A single group of randomly distributed oscillators and the remaining oscillators in a coherent state (Figs. 11(c) and 12(c)). 4) Coherently evolving network (Figs. 11(d) and 12(d)). 
Then, a study of different dynamical states can be carried out by defining the standard deviation
\begin{equation}
\sigma_{x}=\Big<\noindent \sqrt{\frac{1}{N}\sum_{i=1}^{N}[x_{i}-<x>]^2} \hspace{0.1cm} \Big>_{t} ,  \hspace{0.1cm} <x>= \frac{1}{N}\sum_{i=1}^{N}x_{i}(t) \notag
\end{equation}
Fig. 13 presents $\sigma_{x}$ as  a function of $\epsilon$. It is apparent that from this plot one cannot make a very clear distinction between different states (incoherent, coherent, and chimera) of coupled systems connected by non-local coupling.
A comparison of Fig. 13 with Figs. 2 and 3 clearly reveals the significance of the transformed variables $z_{i}$.

\section*{Appendix B: Identification of cluster states}
In the study of coupled systems with nonlocal interaction, at the transition towards a coherent state, we also obtain cluster states for certain values of $\epsilon$. When this states occurs, the smooth profile of the coherent state breaks up into two or three parts.

As an example, we consider a system of nonlocally coupled Lorenz systems with $N=500$, and $\epsilon=8.50$. The snapshots and space-time plots of $x_{1,i}$ and $z_{1,i}$ are shown in Fig. 14. Fig. 14(a) indicates that the smooth profile structure breaks and a few $x_{i}$ values  deviate from the profile. This indicates that a cluster state exists in the coupled system and the corresponding space-time plots of  $x_{i}$ (Fig. 14(b)) also confirms the existence of a cluster state.

In our present study  we identify the existence of a cluster state irrespective of the $\epsilon$ value, if the following condition is satisfied:
\begin{eqnarray}
...\approx z_{i-2}\approx z_{i-1}\approx z_{i}, z_{i}\neq z_{i+1},  \notag  \\
 z_{i+1} \neq z_{i+2}, z_{i+2}\approx z_{i+3} \approx z_{i+4} \approx...  ~ \forall ~~  t  \notag
\end{eqnarray} 
The above definition corresponds to a discontinuity in the values of the variable $z$ at the point $i$. 

Figures 14(c) and 14(d) show the existence of clusters which satisfy the above condition. In Figs.  14(c) and (d) the deviated values of $z_{1,i}$ are removed by the concept of removable discontinuity~\cite{dis1992} for the calculation of $\sigma_{l}$ and $S$.  Figs. 14(e) and 14(f) depict the snapshot/space-time plots of $z_{1,i}$ (after removing the deviated values of  $z_{1,i}$).

\begin{figure}
\centering
\includegraphics[width=1.0\columnwidth=0.9]{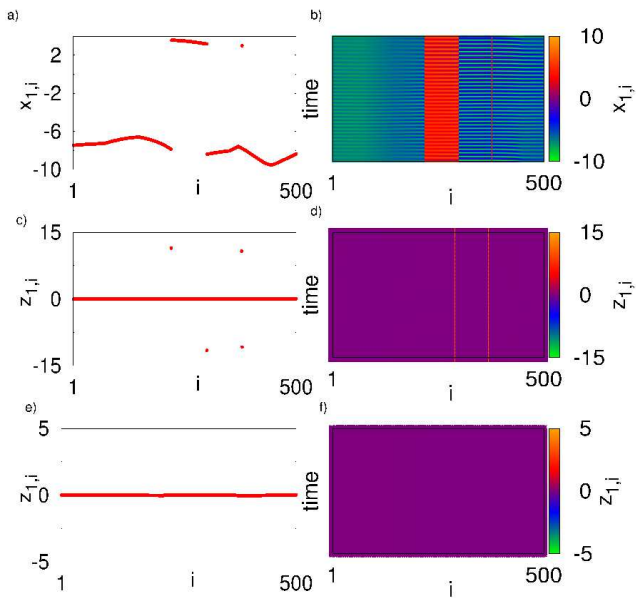}
\caption{\label{fig14}(color online) Snapshots of variables (a) $x_{1,i}$ (c) $z_{1,i}$ (e) $z_{1,i}$ (after removing deviated points by the concept removable discontinuity), space-time plots (b) $x_{1,i}$ (d) $z_{1,i}$ (f) $z_{1,i}$ (after removing deviated points) in the coupled Lorenz system with $N=500$, $r=0.3$ and $\epsilon=8.50$.}
\end{figure}
\end{appendix}

\end{document}